\documentclass[a4paper,12pt]{article}
\usepackage[utf8]{in
    putenc}
\usepackage{cancel}
\usepackage{ulem}
\usepackage{amsfonts}
\usepackage{amssymb}
\usepackage{graphicx}
\usepackage{amsmath}
\usepackage{enumerate}
\usepackage{mathtools}
\usepackage{subfig}
\usepackage{color}
\usepackage{float}
\usepackage{here}
\usepackage{cite}
\usepackage{mathrsfs}
\usepackage{float,epsfig}
\usepackage{dcolumn}

\usepackage{graphicx}
\usepackage{bm}
\usepackage{amsmath,amssymb,amsthm}
\usepackage[colorlinks=true,linkcolor=blue,citecolor=red]{hyperref}
\makeatother

\begin{document}

\title{\bf \Large 
Entanglement Entropy and Phase Portrait   of f(R)-AdS Black Holes
in the Grand Canonical Ensemble
}

\author{
A. Belhaj$^{1}$\footnote{belhaj@unizar.es }, H. El Moumni$^{2,3}$\footnote{hasan.elmoumni@edu.uca.ma (Corresponding author)}\\
}

\maketitle


\vspace{-3.6em}

\begin{center}

\textit{$^a$
ERPTM,  Polydisciplinary Faculty, Sultan Moulay Slimane
University,
 B\'eni Mellal, Morocco.}\\ [0.5em]

\textit{$^b$
EPTHE, Faculty of Sciences, Ibn Zohr University, B.P 8106, Agadir, Morocco.}\\ [0.5em]

\textit{$^c$
High Energy Physics and Astrophysics Laboratory, Faculty of Science Semlalia, Cadi Ayyad University, 40000 Marrakesh, Morocco.}\\ [0.5em]
\end{center}

\vspace{1em}

\begin{abstract}
{\noindent}
In this   work,  we investigate the thermodynamical
   behavior  of  charged AdS black holes from $f(R)$ gravity  corrections with a
constant Ricci scalar curvature in  the grand-canonical ensemble.
Using the  holographic entanglement entropy,  we  show that the
physical observables  behave     as in  the case of the thermal
entropy.  By performing numerical computations associated with the
thermodynamical   quantities versus the entanglement entropy, we
confirm that the same phase portrait persists  in the holographic
framework. In the grand-canonical ensemble, the present result
supports the former finding which reveals that the  charged $f(R)$
AdS black holes behave much like RN-AdS black holes.
\end{abstract}
\newpage
\tableofcontents

\section{Introduction}

Recently,  the  $f(R)$ gravity has received  a remarkable attention
in connections with many  theoretical  physics  subjects. It has
been approached from different angles including gravitation and
cosmology associated with high energy physics problems. In
particular, it has been used to understand, among others,  the
history of universe, especially the
 current cosmic accelerations,  the inflation and  the structure formation in the early Universe \cite{FR2,FR3}.
 It has been
observed that the  black holes in such backgrounds  bring  a
different physics with respect  of the one based on Einstein gravity
by introducing  either higher powers
 of the scalar curvature $R$ or  the Riemann and  the Ricci tensors, including their  derivatives to the lagrangian  contributions \cite{FR1,FR2,FR3,FR4,FR5}.

More recently,  the notion  of the  extended phase space
 comes up with a  new vitality into the study of the black hole thermodynamics
 via the identification of the cosmological constant with the pressure
   and its conjugate quantity  with  the thermodynamic volume
   \cite{Kastor}. Concretely,
 it has been shown that black
holes are, in general,  quite analogous to Van der Waals fluids
\cite{KM,our} leading to a remarkable result. Many extensions  of
these works have been elaborated for rotating  and hairy black holes
\cite{our1,our2}, including the ones embedded in  high curvature
theories of superstrings and M-branes \cite{our3,our4,our5,our6}.
More exotic results as holographic heat engine \cite{our7} as well
as other technics ranging
 from the behavior of the quasi-normal modes \cite{our8,our9}, microscopic structure \cite{simo2} and Joule-Thomson expansion \cite{simo1} 
 to chaos structure
 \cite{Chabab:2018lzf} have consolidated the similarity with the Van der Waals fluids.
On the other hand, another  beautiful analogy is the probe of
 the critical behaviors of  the AdS black holes using the AdS/CFT
 tools including
   entanglement  entropy, Wilson loop  and two point correlation
   functions
   \cite{ch12,ch13,ch14,holo1,holo2,holo3,holo4,XXX,ch15,X3}.

It turns out that the thermodynamics of  the $f(R)$ gravity attracts
a huge part of attention in recent literature. In particular, the
four dimensional charged black holes in the $R+f(R)$ gravity  have
been elaborated in the canonical ensemble context \cite{Chen}.
Concretely, the phase transition and its thermodynamic geometry have
been extensively investigated\cite{xiong6}. These results have been
extended to the grand-canonical ensemble\cite{base}. This extension
has shown a remarkable difference in the associated black hole
physics in both ensembles. In the case of  four dimensional charged
AdS blacks in the $R+f(R)$ gravity with  a constant curvature in the
grand canonical ensemble, the phase transition has been investigated
in some details. Precisely,  it  has been revealed that the
corresponding thermodynamics is quite different from the one
appearing in  the canonical ensemble. For fixed electric potentials,
other physical quantities have been also obtained  including the
specific heat and the isothermal compressibility coefficient showing
certain divergences  in the associated expressions. Moreover, the
thermodynamic geometry has been also   discussed to deal with the
corresponding phase structure.

In this  paper,
  we  holographically  investigate  the thermodynamical  behavior of AdS black holes using  the entanglement entropy in the grand-canonical ensemble.
   Concretely, we discuss the thermodynamical  aspects
of
 charged $f(R)$ AdS black holes with a constant Ricci scalar curvature
 in such an   ensemble.  Calculating  the holographic entropy, we
 show that the physical observable behave like in the thermal
 entropy case. Plotting  thermodynamical
 quantities versus the entanglement entropy, we confirm that the
 same phase portrait persists in the holographic framework. In the
 grand canonical ensemble, the present results support the former
 finding revealing  that the charged $f(R)$ AdS black holes behave
  like RN-AdS black
holes.

The organization of the manuscript is as follow. In section 2, we
give a concise review on  the  phase transition of  the $f(R)$ AdS
black holes in the  grand-canonical ensemble. Section  3 concerns
the holographic entanglement entropy and the phase structure  of
such black hole solutions. In section 4, we elaborate  the
geometrothermodynamics for  the $f(R)$ AdS black holes in  order to
discuss the  corresponding phase structure. The last section is
devoted to conclusions and remarks.
\section{Phase transition of
$f(R)$ AdS black holes in grand-canonical ensemble}\label{sec:2} In
this section, we give a concise review on the  black hole solutions
in the $R+f(R)$ gravity with a constant curvature. The action
describing a four-dimensional charged AdS black hole solution  in
such  gravity backgrounds is given by \cite{fBH6}
\begin{eqnarray}
\mathcal{I}=\int_{\mathcal{M}} d^{4}x\sqrt{-g}[R+f(R)-F_{\mu\nu}F^{\mu\nu}].
\label{action}
\end{eqnarray}
Here, $R$ denotes the Ricci scalar curvature while $f(R)$ is an
arbitrary function of $R$. It is recalled that $F_{\mu\nu}$  is the
abelian electromagnetic field tensor being  related to the
electromagnetic potential $A_\mu$ as follows
\begin{equation}
F_{\mu\nu}=\partial_{\mu}A_{\nu}-\partial_{\nu}\partial_{\mu}.
\end{equation}
To get the corresponding equations of  motion, one may  use the
variation of the action (\ref{action})  with respect to these
fields. For the  gravitational field $g_{\mu\nu}$, one gets
\begin{eqnarray}
R_{\mu\nu}[1+f'(R)]-\frac{1}{2}g_{\mu\nu}[R+f(R)]+(g_{\mu\nu}\nabla^2-
\nabla_{\mu}\nabla_{\nu})f'(R)=T_{\mu\nu}, \label{Esteq}
\end{eqnarray}
while for  the gauge field $A_{\mu}$,  one obtains
\begin{eqnarray}
\partial_{\mu}(\sqrt{-g}F^{\mu\nu})=0.
\label{Emeq}
\end{eqnarray}

 Moon \textit{et al }\cite{fBH6} consider the simple
case where   the Ricci scalar curvature is constant  $R=R_0=const$
which provides an analytical solution of the equation (\ref{Esteq}).
Under the constant Ricci scalar curvature assumption,  this equation
becomes
\begin{eqnarray}
R_{\mu\nu}[1+f'(R_0)]-\frac{g_{\mu\nu}}{4}R_0[1+f'(R_0)]
=T_{\mu\nu}. \label{Esteq1}
\end{eqnarray}
In what follows, we discuss the black hole in such a concrete
geometry.

\subsection{Black holes in the $R+f(R)$ gravity with constant curvature}
It turns out that a  four-dimensional charged AdS black hole
solution in the $R+f(R)$ gravity with constant curvature  has been
obtained with its thermodynamic quantities. In particular,  energy,
entropy, heat capacity and Helmholtz free energy  have been computed
and  discussed  ~\cite{fBH6}. Moreover, the $P-V$ criticality of
this solution has been  investigated in ~\cite{Chen}. The
coexistence curve and the density  number  of  molecules for such a
black hole solution have  been dealt with  in ~\cite{xiong5}.
Concretely, it has been studied the  phase transition in the
canonical ensemble for a black hole having   fixed
charges\cite{xiong6}.\\
 To  understand such activities,   consider the RN-AdS  black hole ~\cite{fBH6}. In this way, the
 metric  line element,  in the $f(R)$ gravity backgrounds,  reads as
\begin{equation}
ds^2=-N(r)dt^2+\frac{dr^2}{N(r)}+r^2(d\theta^2+sin^2\theta
d\phi^2),\label{1}
\end{equation}%
where the $N(r)$ function  should satisfy Eq.\eqref{Esteq1}  and
takes the following form
\begin{eqnarray}
N(r)&=&1-\frac{2m}{r}+\frac{q^2}{br^2}-\frac{R_0}{12}r^2\label{2}.
\end{eqnarray}
Here, the $b$ quantity is written as follows
\begin{eqnarray}
b&=&1+f'(R_0).\label{3}
\end{eqnarray}%
In this solution, one has  $b>0$ and $R_0<0$. Taking $b=1$ and
$R_0=-12/\ell^2=4\Lambda$,  it is observed that  such a black hole
solution can be reduced to the  RN-AdS black hole. Moreover,  it has
been shown that the  black hole ADM mass $M$ and the electric charge
$Q$ are related to the parameters $m$ and $q$,
respectively~\cite{fBH6}. They are given by
\begin{equation}
M=mb,\;\;\; Q=\frac{q}{\sqrt{b}}.\label{4}
\end{equation}%
According to ~\cite{Chen},  the corresponding  thermodynamic
quantities can be derived  using the associated computations. Some
of them are given by the following expressions
\begin{eqnarray}
T&=&\frac{N'(r_+)}{4\pi}=\frac{1}{4\pi
r_+}(1-\frac{q^2}{br_+^2}-\frac{R_0r_+^2}{4})\label{6}
\\
S&=&\pi r_+^2b\label{7}
\\
\Phi&=&\frac{\sqrt{b}q}{r_+}\label{8}
\end{eqnarray}
 where $T$, $S$ and $\Phi$  indicate  the Hawking temperature, the entropy and
the electric potential respectively.  It is recalled that the
entropy can be obtained  by exploiting  the  Wald method developed
in \cite{FR3}. More details on  such calculations   can be
found in \cite{Wald}.\\
Having discussed the  thermodynamic quantities on such gravity
backgrounds, we move now to investigate the associated  phase
transition.

\subsection{Phase transition of $R+f(R)$ AdS black holes}
\label{sec:3} For facility reasons, it is useful to rebuilt the
important quantities in terms of the entropy $S$
 and the electric potential $\Phi$. Indeed,  the calculations show that the Hawking
 temperature  reads as

\begin{equation}
T=T(S,\Phi)= \frac{4b^2\pi-bR_0
S-4\pi\Phi^2}{16\pi^{3/2}b^{3/2}\sqrt{S}}.\label{9}
\end{equation}
Using this equation,  one can  obtain the following derivative
relations, which will be used later,
\begin{eqnarray}
\left(\frac{\partial T}{\partial
S}\right)_\Phi&=&\frac{-b(4b\pi+R_0S)+4\pi\Phi^2}{32\pi^{3/2}(bS)^{3/2}}
,\label{10}
\\
\left(\frac{\partial^2 T}{\partial
S^2}\right)_\Phi&=&\frac{12b^2\pi+bR_0
S-12\pi\Phi^2}{64\pi^{3/2}b^{3/2}S^{5/2}} .\label{11}
\end{eqnarray}
The solution of the first equation  $\left(\frac{\partial
T}{\partial S}\right)_\Phi=0$  is  given by
\begin{equation}
S_1=\frac{-4\pi(b^2-\Phi^2)}{bR_0}.\label{12}
\end{equation}
Taking   $b>0$ and $R_0<0$, the constraint $0<\Phi<b$  is needed to
ensure   the  positivity  of  the entropy appearing  in
Eq.(\ref{12}).
 Substituting Eq.(\ref{12}) into Eq.(\ref{11}),  one gets
\begin{equation}
\left(\frac{\partial^2 T}{\partial
S^2}\right)_\Phi\mid_{S=S_1}=\frac{bR_0^4}{256\pi^3[R_0(-b^2+\Phi^2)]^{3/2}}>0.\label{13}
\end{equation}
The Hawking temperature for both  cases $0<\Phi<b$ and $\Phi>b$ is
illustrated in Fig.\textcolor{blue}{1a} and Fig.\textcolor{blue}{1b} respectively.

 It follows
from Fig.\textcolor{blue}{1a} that  there exists a  minimum temperature
associated with the condition $0<\Phi<b$. Substituting Eq.(\ref{12})
into Eq.(\ref{9}),  this  temperature  is  explicitly given by
\begin{equation}
T_{min}=\frac{\sqrt{-R_0(b^2-\Phi^2)}}{4b\pi} .\label{14}
\end{equation}
It is  observed, however, that   the Hawking temperature increases
monotonically when $\Phi>b$. This behavior is  represented  in
Fig.\textcolor{blue}{1b}.


\begin{figure}[!ht]
\begin{center}
\includegraphics[width=8cm]{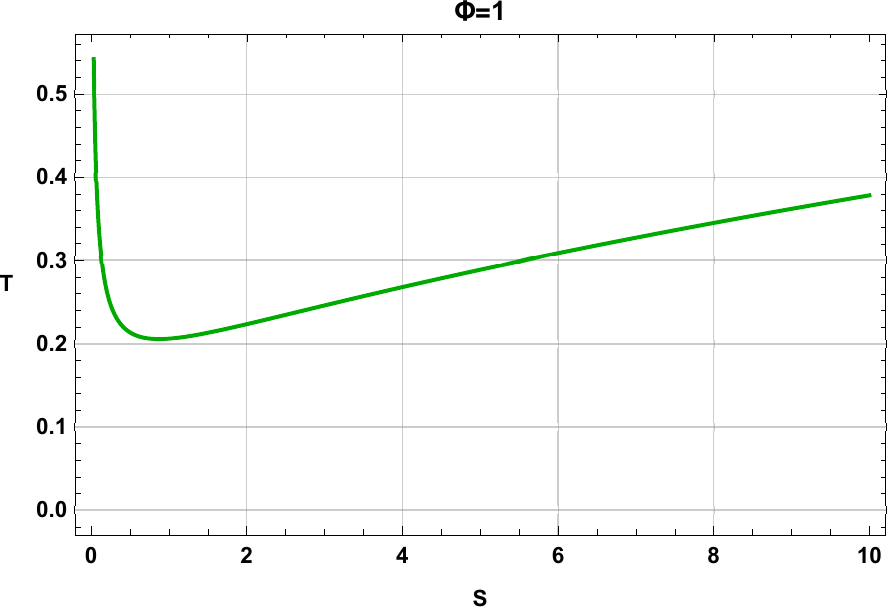}\vspace{1cm}\includegraphics[width=8cm]{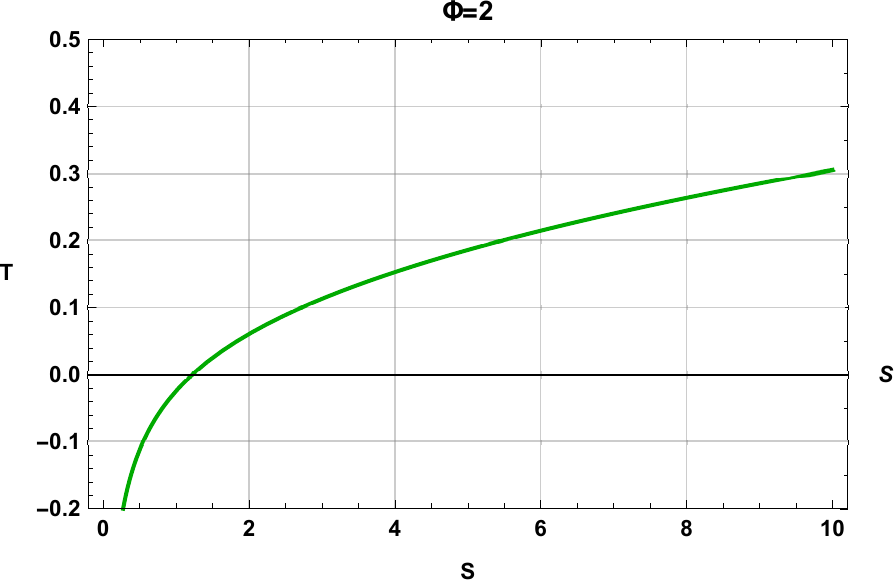}
\vspace{-0.8 cm}
\caption{\footnotesize The behavior  of the Hawking temperature in function
of the entropy for different values of the electric potential. }\label{fig1}
\end{center}
\end{figure}


To approach the phase transition, one may examine the specific heat
behavior for fixed charges. In the grand-canonical ensemble
associated with  a fixed electric potential of $f(R)$ AdS black
hole, the specific
 heat takes the following form
\begin{equation}
C_\Phi=T(\frac{\partial S}{\partial T})_\Phi=\frac{2S(-4b^2\pi+bR_0
S+4\pi \Phi^2)}{4b^2\pi+bR_0 S-4\pi \Phi^2}.\label{15}
\end{equation}
It is remarked   that  the denominator of Eq.(\ref{15}) is exactly
the same as the numerator of Eq.(\ref{10}). This shows   that the
divergence of $C_\Phi$  corresponds to  the minimum Hawking
temperature.

 In this way, it is  easy  to  obtain  the condition
showing  the  divergence of the heat capacity $C_\Phi$. Indeed, it
is given by the following constraint
\begin{equation}
4 \pi  b^2+b R_0 S-4 \pi  \Phi ^2=0.\label{19}
\end{equation}
This condition  can be  solved  by taking
\begin{equation}
S=\frac{4 \pi \left( b^2-  \Phi ^2\right)}{b R_0}.\label{20}
\end{equation}
Using the restrictions $b>0$ and $ R_0<0$, the above root  is
accepted physically only  for  $0<\Phi<b$.

 The corresponding  behaviors are plotted in  Fig.\textcolor{blue}{2a} and Fig.\textcolor{blue}{2b}.
 In particular,  Fig.\textcolor{blue}{2a} is associated with  the case of $0<\Phi<b$ while Fig.\textcolor{blue}{2b}
corresponds to the  case of $\Phi>b$. It follows form these
computations  that the specific heat $C_\Phi$ involves  a divergence
for $0<\Phi<b$. However, the divergence  is removed in the case of
$\Phi>b$. This result differs from  the one obtained  in the
canonical ensemble~\cite{xiong6}, where the system can  involve two,
one or no divergence points for the specific heat $C_Q$.


\begin{figure}[!ht]
\begin{center}
\includegraphics[width=8cm]{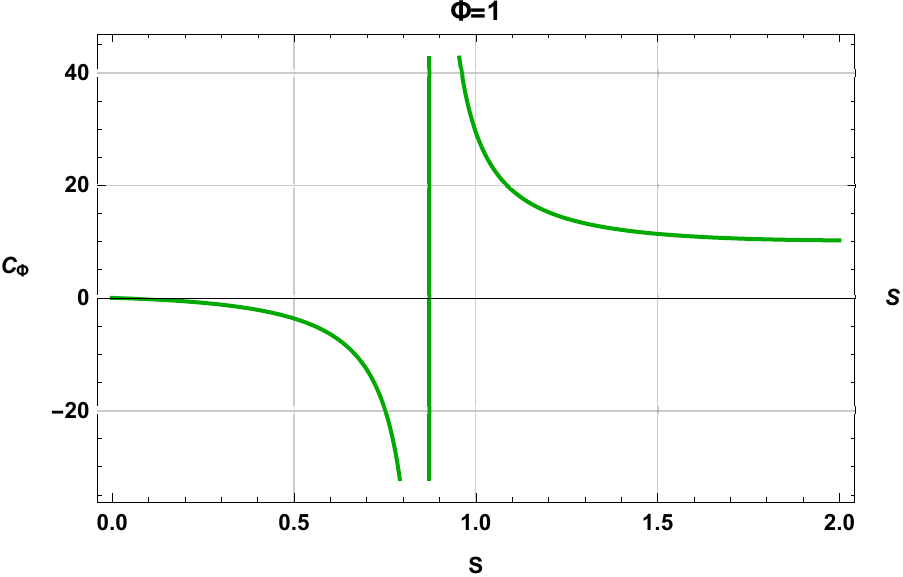}\vspace{1cm}\includegraphics[width=8cm]{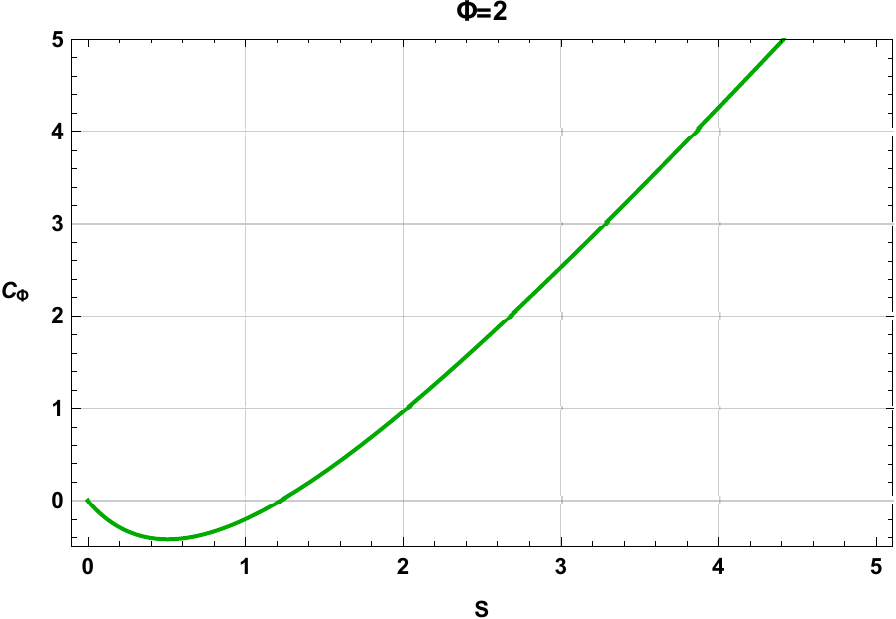}
\vspace{-0.8 cm}
\caption{\footnotesize The behavior  of the heat Capacity in function of
the entropy for different values of the electric potential. }\label{fig2}
\end{center}
\end{figure}



\section{Holographic entanglement entropy and the phase structure\label{section3}}

Having understood the essential of thermodynamical behaviors of such
black holes in the grand-canonical ensemble, and motivated by recent
research  in  holographic domain \cite{ch12,ch13,ch14,holo1,holo2},
we would like to employ the non local observables like the
holographic entanglement entropy to check  whether  this quantity
can respect  the grand-canonical ensemble phase structure. To do so,
let us consider a  quantum  field theory described by a density
matrix $\rho$, with  $A$ is  a  region of a   spacetime Cauchy
surface and $A^{c}$ is its complement. In this way, the entanglement
entropy between these two regions can be defined  as
\begin{equation}
S_{A} = -\mathrm{Tr}_{A}{(\rho_{A}\log{\rho_{A}})},
\end{equation}
where $\rho_{A}$ is the reduced density matrix of $A$ given by
\begin{equation}\rho_{A}=\mathrm{Tr}_{A^{c}}{(\rho)}. \end{equation}

 According to \cite{dong,camps}, it is well known
that when  a  gravitational theory contains higher powers of the
curvature,   the Ryu and Takayanagi formula \cite{7,8}
 must include additional contributions originated  form
the extrinsic curvature. Generally, these  corrections read as
\begin{equation}\label{general}
S_A=2\pi\int d^d x\sqrt{g}\left\{\frac{\partial L}{\partial R_{z\bar{z}z\bar{z}}}+\sum_{\alpha}
\left(\frac{\partial^2 L}{\partial R_{zizj}\partial R_{\bar{z}k\bar{z}l}}\right)\frac{8 \mathcal{K}_{zij}\mathcal{K}_{\bar{z}kl}}{q_\alpha}
\right\}
\end{equation}
where $L(R_{\mu\rho\nu\sigma})$ is the associated  Lagrangian and
$\mathcal{K}$ indicates  the extrinsic curvature. It is recalled
that $\alpha$ and $q_\alpha$ stand for the a  term of expansion and
an anomaly term, respectively, while the couple  $(z,\bar{z})$
denotes  the orthogonal complex coordinates \footnote{More details
can be found in  the appendix of \cite{dong}.}.  In the   present
gravity model, the Lagrangian $L$ depends on the Riemann tensor only
through the Ricci scalar $R$. Under this assumption,  the general
formula given in \eqref{general} reduces to
\begin{equation}\label{simple}
S_A=-4\pi \int d^2 x\sqrt{g}\frac{\partial L}{\partial R}
\end{equation}
The second term in \eqref{general},  which involves extrinsic
curvatures, vanishes  due to the fact that  $R$ does not contain
components of the form $R_{zizj}$.  It has been shown that this is a
 consistency verification of the formula. Indeed, transforming $f(R)$
gravity to a theory of Einstein gravity coupled to a scalar, and
using the Ryu-Takayanagi formula,  we can  find the same
entanglement entropy \cite{dong,camps}.  Precisely,  the
entanglement entropy takes the following form
\begin{equation}
S_{A} = \frac{\text{Area}(\Gamma_A)}{4 G_N},
\end{equation}
where $\Gamma_A$ is a codimension-2 minimal surface with the
boundary condition $\partial \Gamma_A=\partial A$, and where
$G_{N}$ is the gravitational Newton's constant.

  For  the present    black hole configurations,  we choose the region $A$ to be a spherical cap on the boundary delimited
   by $\theta\leq\theta_0$ and parameterized by the coordinate
   $r(\theta)$. Thus,   the minimal area  corresponding to  the entanglement entropy can
be written as
\begin{equation}\label{AA}
\mathcal{A}  = 
 2\pi\int_{0}^{\theta_{0}} r^{2}\sin^{2}{\theta}\sqrt{\frac{(r')^{2}}{f{(r)}}+r^{2}} d\theta\,,
\end{equation}
where $r' \equiv \frac{dr}{d\theta}$ and $\theta_0$ is the boundary
of the entangled region. The function $r{(\theta)}$ can be
obtained by interpreting  Eq.(\ref{AA}) as a Lagrangian and  solving
the following equation of motion
\begin{eqnarray}\nonumber
0&=&r'(\theta )^2 [\sin \theta  r(\theta )^2 f'(r)-2 \cos \theta
r'(\theta )]-2 r(\theta ) f(r) [r(\theta ) (\sin \theta  r''(\theta
) +\cos \theta  r'(\theta )) -3 \sin \theta  r'(\theta )^2]\\ &+&4
\sin (\theta ) r(\theta )^3 f(r)^2\label{sysEE}
\end{eqnarray}
with the following boundary conditions
\begin{equation}\label{bc}
 r'(\theta)=0,\;\; r(0)=r_0.
\end{equation}

To  regularize the  entanglement entropy,  we subtract  the area of
the minimal surface in  the  pure AdS geometry  whose  the boundary
is also $\theta=\theta_{0}$ with
\begin{equation}
r_{AdS}{(\theta)} =
L\left(\left(\frac{\cos{\theta}}{\cos{\theta_{0}}}\right)^{2}-1\right)^{-1/2}\,.
\end{equation}

Performing  numerical calculations, we can easily
 solve the  equation  Eq.\eqref{sysEE} by using  the boundary conditions of
Eq.\eqref{bc} and taking
 $\theta_{0} = 0.2$  with different electric potential values  namely $\Phi=1$ and  $2$, while the parameter $b$ is set to be $1.5$.
 In the present investigation, the
Ultra Violet cutoff is    chosen    to be $\theta_{c} =  0.199$.

\begin{figure}[!ht]
\begin{center}
\includegraphics[width=8cm]{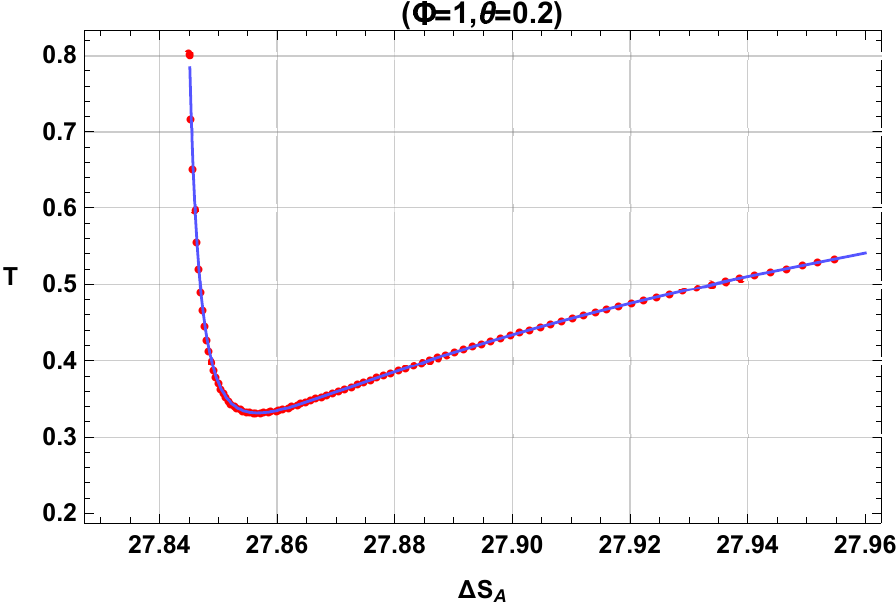}\vspace{1cm}\includegraphics[width=8cm]{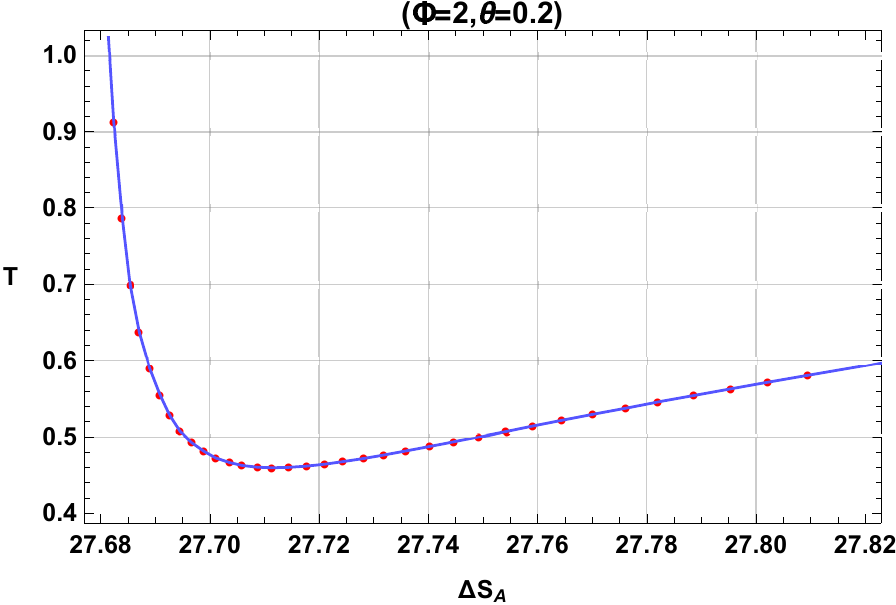}\\
\includegraphics[width=8cm]{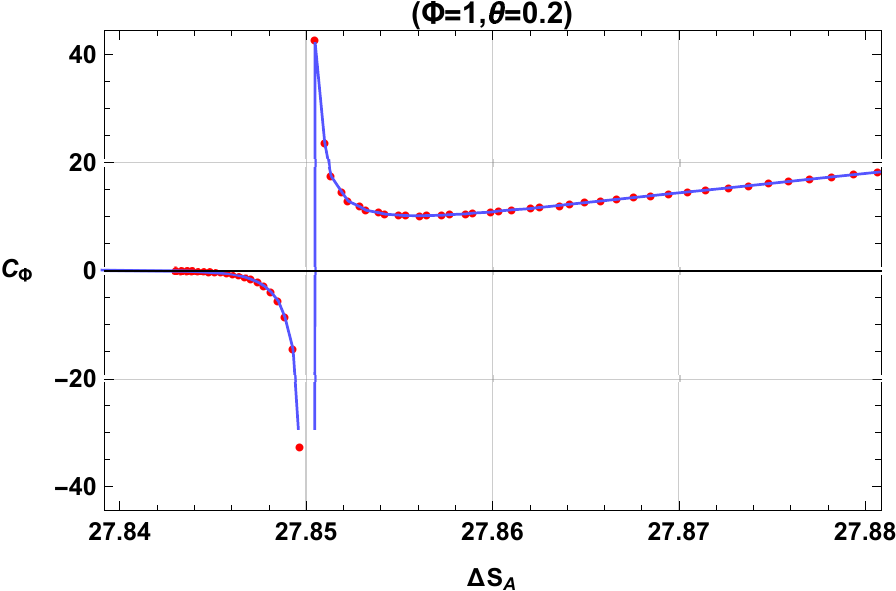}\vspace{1cm}\includegraphics[width=8cm]{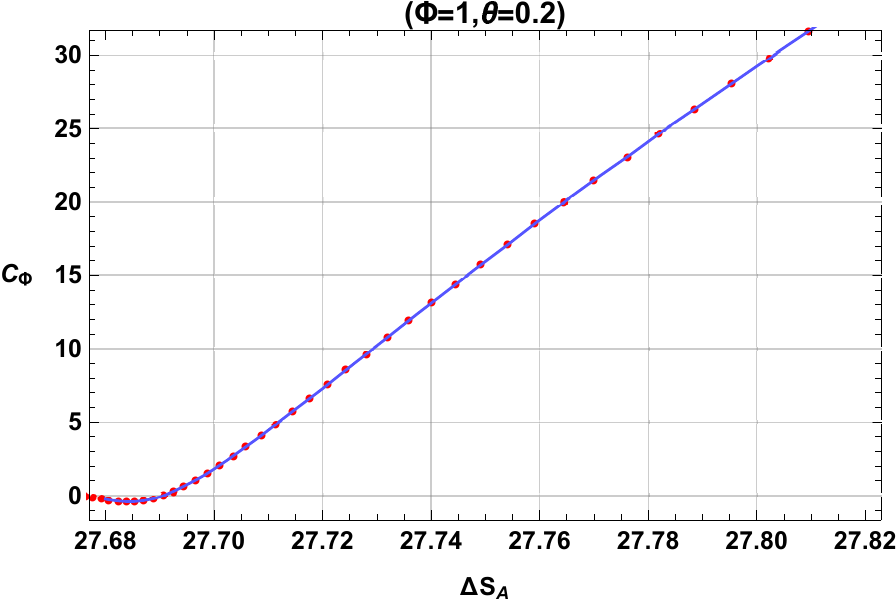}
\vspace{-0.8 cm}
\caption{\footnotesize The behavior  of the temperature and the heat Capacity in function of
the entanglement entropy for different values of the electric potential.
 We have also show the data points which were used to create the interpolation (red points). }\label{fig3}
\end{center}
\end{figure}

Now  we are in position  to give a comparative analysis  with the
thermal structure. For this reason,  we plot in Fig.\ref{fig3} the
relations between the Hawking temperature and the heat capacity
versus
 the holographic
entanglement entropy  for different values of parameters the
electric potential  $\Phi$ with the chosen $\theta_{0}$.

%
%

It is remarked  that  Fig.\ref{fig3} shares similarities with
Fig.\ref{fig1} and Fig.\ref{fig2}. Especially,  it has been observed
a minimum of the temperature in the case $0<\Phi<b$ and a  monotony
when $\Phi>b$. For the heat capacity also the divergence point is
recovered in the first case as the continuity in the second one.
Having provided the same behavior for the both plans $T-S$ and
$T-\Delta S_A$, we  can conclude that the phase structure of the
holographic entanglement entropy is the same as in  the thermal
entropy structure  in   the grand-canonical ensemble framework.

\section{Thermodynamic geometry in grand-canonical ensemble}
To completely  explore this phase transition and reinforce the
conclusion of the previous section, we will recall the
thermodynamical geometry tools, such as  the Weinhold geometry
\cite{cwein} and  the Ruppeiner one
 \cite{c14}.

The Weinhold metric  is defined as the second derivative
 of the internal energy with respect to the entropy and other extensive quantities in the energy representation,
  while the Ruppeiner metric  is related to  the Weinhold metric by a temperature conformal
  scale factor. Indeed, one has  the following relation
 \begin{equation}\label{3.5}
\;\;\;\;\;\; ds_R^2=\frac{1}{T} ds_W^2.
 \end{equation}
In this  context,  we can evaluate the thermodynamical curvature of
the  presented black holes.  Concretely, the Weinhold and the
Ruppeiner metric are given, respectively,   by
\begin{equation}\label{3.8}
g^W=\left(\begin{array}{cc}M_{SS} & M_{S \Phi} \\M_{\Phi S} & M_{\Phi \Phi}\end{array}\right),\quad  g^R=\frac{1}{T}\left(\begin{array}{cc}M_{SS} & M_{S\Phi} \\M_{\Phi S} & M_{\Phi\Phi}\end{array}\right),
\end{equation}
where $M_{ij}$ stands for $\partial^2 M/ \partial x^i\partial x^j$,
 $x^1=S$ and $x^2=\Phi$.
 According to \cite{base}, the expressions of Weinhold  and  Ruppeiner scalar curvatures can be obtained explicitly.  Precisely, they are  given,
 respectively, by
\begin{equation}\label{rw}
R^W=\frac{8 \pi ^2 b^{5/2} \left(8 \pi  b^3 R_0 \sqrt{\frac{S}{b}}+32 \pi ^{3/2} b^3+3
   b^2 R_0^2 S \sqrt{\frac{S}{b}}+12 \sqrt{\pi } b^2 R_0 S-96 \pi ^{3/2} b
   \Phi ^2+4 \sqrt{\pi } R_0 S \Phi ^2\right)}{\sqrt{S} \left(4 \pi  b^2+b
   R_0 S-4 \pi  \Phi ^2\right)^3},
   \end{equation}
   \begin{equation}\label{rr}
   R^R= \frac{\pi ^4 b\ \mathcal{A}(S)}{S \left(4 \pi  b^2+b R_0 S-4 \pi  \Phi ^2\right)^3
   \left(-4 \pi  b^2+b R_0 S+4 \pi  \Phi ^2\right)}
   \end{equation}
  where the function $\mathcal{A}(S)$ stands for
 \begin{eqnarray}\nonumber
 \mathcal{A}(S)&=&\frac{1}{\pi ^5}[-64 \pi ^3 b^5 R_0^2 S^2-768 \pi ^5 b^5 \Phi ^2-112 \pi ^2 b^4 R_0^3
   S^3-320 \pi ^4 b^4 R_0 S \Phi ^2-60 \pi  b^3 R_0^4 S^4\\
   \nonumber&+&336 \pi ^3 b^3
   R_0^2 S^2 \Phi ^2-512 \pi ^5 b^3 \Phi ^4-9 b^2 R_0^5 S^5+36 \pi ^2 b^2
   R_0^3 S^3 \Phi ^2-128 \pi ^4 b^2 R_0 S \Phi ^4\\&+&16 \pi ^3 b R_0^2 S^2
   \Phi ^4+1280 \pi ^5 b \Phi ^6+448 \pi ^4 R_0 S \Phi ^6]
 \end{eqnarray}

\begin{figure}[!ht]
\begin{center}
\includegraphics[width=8cm]{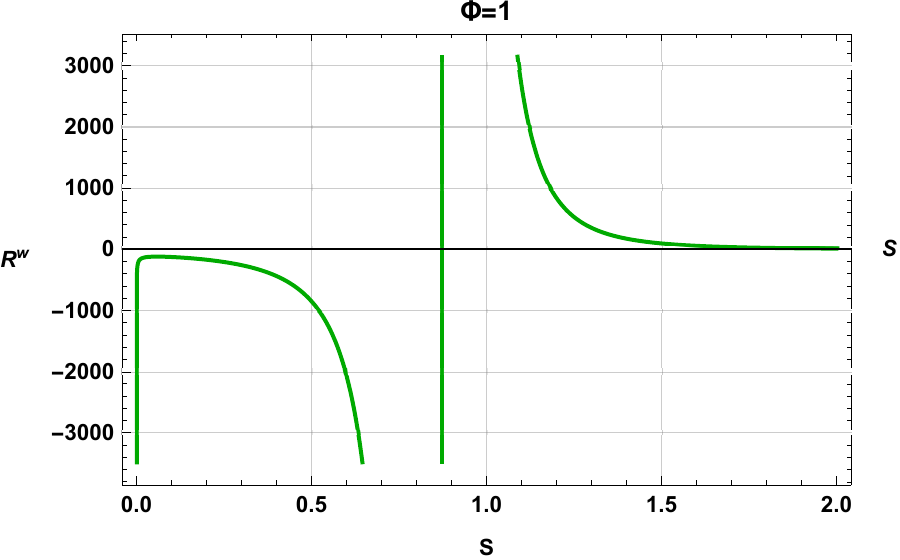}\vspace{1cm}\includegraphics[width=8cm]{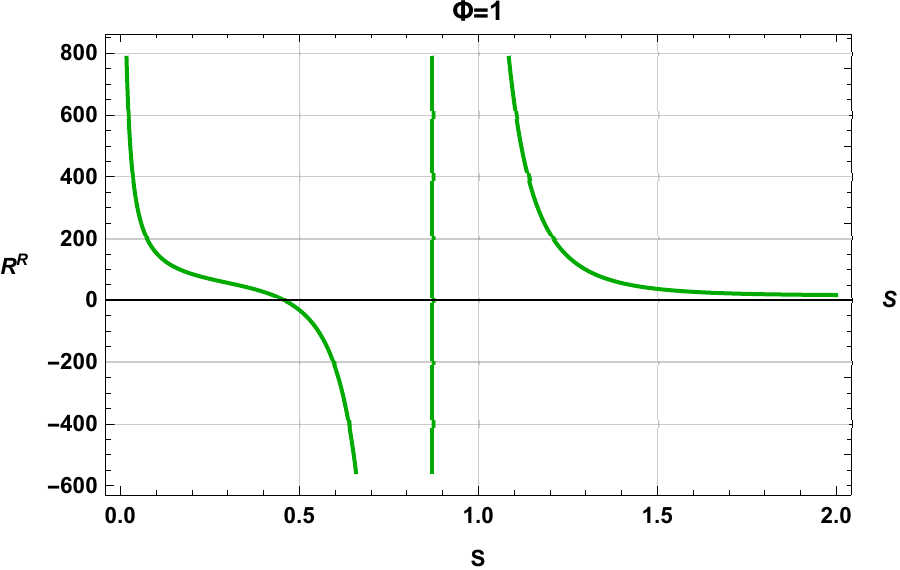}
\vspace{-0.8 cm}
\caption{\footnotesize The behavior  of scalar curvatures in function of
the entropy for  $b=1.5$ and  $\Phi=1$. }\label{fig4}
\end{center}
\end{figure}


 Comparing Eq.\eqref{rw} and Eq.\eqref{rr} with Eq.\eqref{19}, one may find
 that the
Weinhold scalar curvature shares the same factor $4 \pi  b^2+b R_0
S-4 \pi  \Phi ^2$ in the denominator as the specific heat does. This
means that  it would diverge exactly where the specific heat
diverges. This is also shown intuitively in Fig.\ref{fig4},
representing the behavior of these scalar curvatures as function of
the entropy $S$.

\begin{figure}[!ht]
\begin{center}
\includegraphics[width=8cm]{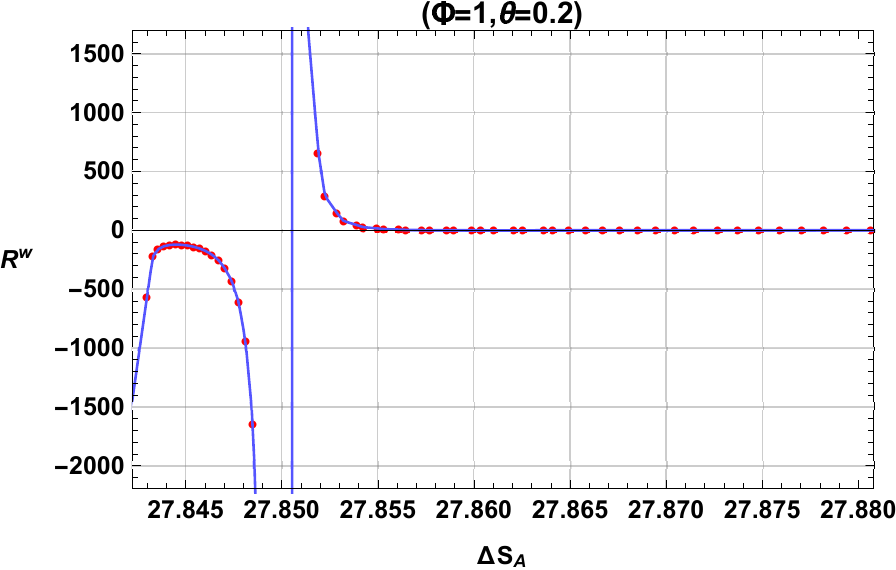}\vspace{1cm}\includegraphics[width=8cm]{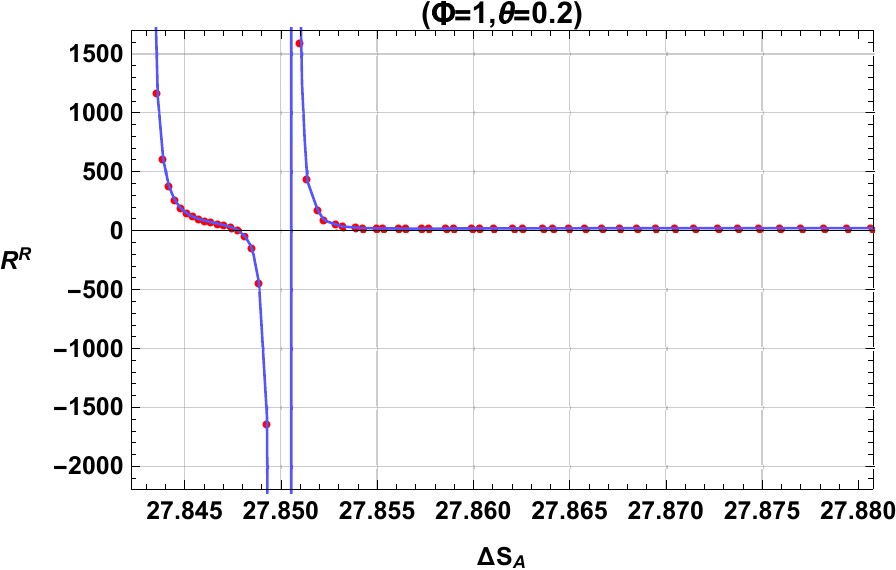}
\vspace{-0.8 cm}
\caption{\footnotesize The behavior  of scalar curvatures in function of
the entanglement entropy for  $b=1.5$ and  $\Phi=1$. Also  data points which were used to create the interpolation are shown (red points). }\label{fig5}
\end{center}
\end{figure}

In the remaining part of this work, we plot also the variation of
the Weinhold and  the Ruppeiner Ricci scalars  in terms of the
holographic entanglement entropy. This is illustrated  in
Fig.\ref{fig5}.

Comparing  with Fig.\ref{fig4}, we can observe  that the holographic
entanglement entropy can  reproduce the same phase portrait  for the
studied
 black hole in the grand-canonical ensemble as the thermal picture.
   A close examination of  the  $T-\Delta S_A$ diagram shows that
   the scalar curvature exhibits  the same singularity  schemes  as the $T-S$ plan  confirming
    the  statement presented   in  the end of Sec.\ref{section3}.

\newpage
\section{Conclusions and discussions}

In this paper,  we have  investigated  the thermodynamic and the
geothermodyanamic
 behaviors in the grand-canonical ensemble by fixing the electric  potential
 of the
  charged AdS black holes in  the $f(R)$ gravity in four dimensions.  Treating  the cosmological
 constant and its  conjugate quantity as thermodynamic variables and
 using
 the holographic entanglement,
this   study  has been  made  within  an extended phase space.
First, we have  presented  the essential  of the thermodynamical
behavior in the thermal structure. Then,  we have moved to  show
that all thermodynamical quantities  exhibit the same behavior
versus
 entropy as well  as the holographic entanglement entropy.
Moreover, we  have found that the phase structure of  such charged
AdS black holes in the grand-canonical
  ensemble can  probe   the  holographic entanglement entropy reproducing  the same thermodynamical
   behavior of the thermal structure.  It has been remarked that  this   provides  a new approach to  understand the phase
   structure
     from the holography point view  associated  with fixed  electric potentials.

     This  work comes up with certain open questions related
     to quantum information theory. In particular, it  would be interesting
      to see if this has any possible connection
with   quantum discord  and complexity properties, by assuming that
Ads black holes can be viewed as    qubit systems. Moreover,   it
should be of relevance to approach quantum information concepts
using quantum gravity in connection with black holes. This will be
addressed elsewhere.


\end{document}